\documentclass{aa}
\usepackage{graphics, epsfig}
\begin{document}

\title{Constraints on the slope of the dark halo mass function by microlensing observables\thanks{In memory of Ruggiero de Ritis, deceased September 2000}}

\author{V. F. Cardone\inst{1,3}
        \and R. de Ritis$^{\dagger}$\inst{2,3}
        \and A. A. Marino \inst{3,4}}

\offprints{V.F. Cardone, \email{winny@na.infn.it}}

\institute{Dipartimento di Fisica E.R. Caianiello, Universit\`a di Salerno, Via S. Allende, 84081 - Baronissi (Salerno), Italy
\and Dipartimento di Scienze Fisiche, Universit\`{a} di Napoli, Complesso Universitario di Monte S. Angelo, Via Cinthia, Edificio N -  80126 Napoli, Italy 
\and Istituto Nazionale di Fisica Nucleare, Sezione di Napoli, Complesso Universitario di Monte S. Angelo, Via Cinthia, Edificio G -  80126 Napoli, Italy 
\and Osservatorio Astronomico di Capodimonte, Via Moiariello, 16 - 80131 Napoli, Italy}
              
\date{Receveid / Accepted }

\abstract{We investigate the dark halo lens mass function (MF) for a wide class of spheroidal non singular isothermal models comparing observed and observable microlensing quantities for MACHO observations towards LMC and taking into account the detection efficiency. We evaluate the microlensing observable quantities, i.e. observable optical depth, number of events and mean duration, for models with homogenous power\,-\,law MF changing the upper and lower mass limits and the flattening of the dark halo. By applying the simple technique of the inverse problem method we are then able to get some interesting constraints on the slope $\alpha$ of the MF and on the dark halo mass fraction $f$ made out by MACHOs consistently with previous results.
}

\titlerunning{The slope of the lens mass function}

\maketitle

\section{Introduction}

The usefulness of gravitational microlensing as a tool to investigate the structure of the galactic dark halo is now firmly established. Since Paczynski's seminal paper (\cite{Pac86}), several collaborations have searched for microlensing events towards LMC (\cite{R97}; \cite{A00}), SMC (\cite{A97c}; \cite{Af98}) and M31 (\cite{Ans97}). While the presence of MACHOs (Massive Astrophysical Compact Halo Objects) has been witnessed by the microlensing events already found, many questions about their nature and the structure of the dark halo are still open.

One of the main quantities that characterise the microlensing properties of a particular MACHO population is the {\it optical depth} $\tau$. This quantity is simply the number of lenses inside the microlensing tube, a cylinder whose axis is the line of sight to the source and with radius equal to the {\it Einstein radius}, defined as (\cite{MR97})
\begin{displaymath}
R_E = \sqrt{\frac{4Gm}{c^2} \frac{D_d (D_s - D_d)}{D_s}} 
= r_E \sqrt{\mu s(1-s)} \ ,
\end{displaymath}
where $m$ is the mass of the lens
\footnote{In this paper we use the terms lens and MACHO as synonymous, even if the lens may be also a visible star.} 
(with $\mu = m/M_{\odot}$), $D_s$ ($D_d$) the distance to the source (the lens), $s=D_d/D_s$ and we have posed $r_E = \sqrt{4GM_{\odot} D_s/c^2}$. In its simplest form, $\tau$ is defined as (\cite{Pac86}; \cite{MR97}; \cite{J98})
\begin{equation}
\tau = \int_0^1{\frac{\pi R_E^2}{m} \rho (s) ds} = 
\frac{4\pi G D_s^2}{c^2}\int_0^1{\rho (s) s (1-s) ds} \ ,
\label{eq: tau}
\end{equation}  
having supposed (as usually) that the halo extends till the source
\footnote{This is a reasonable hypothesis if the source is in Magellanic Clouds, but it is not if the target of observations is M31. However, in this paper we limit ourselves to observations towards LMC only.},
while $\rho(s)$ is the MACHOs mass density. Optical depth is obtained from observations using the formula (\cite{A97b})
\begin{equation}
\tau_{meas} = 
\frac{\pi}{4E} \sum_i 
\frac{t_{E,i}}{\varepsilon(t_{E,i})} \ ,
\label{eq: taumeas}
\end{equation}
being $E$ the total exposure in star\,-\,years (which is simply the number of monitored object multiplied by the total period of observations), $t_E = 2R_E/v_{\perp}$ the {\it Einstein diameter crossing time} (with $v_{\perp}$ the MACHO' s velocity transverse to the line of sight), $\varepsilon(t_{E,i})$ the detection efficiency for the i\,-\,th event and the sum is over the observed events.

Having estimated $\tau_{meas}$ from observations with Eq.(\ref{eq: taumeas}) and calculated the predicted optical depth for an assumed model of full MACHO halo from Eq.(\ref{eq: tau}), an easy way to evaluate the halo mass fraction composed by MACHOs is to compute the ratio between these two quantities to get $f = \tau_{meas}/\tau$. However, Alcock {\it et al.} (1997a) have pointed out that Eq.(\ref{eq: taumeas}) is not a measure of the total optical depth, but only of the optical depth of events which fall in the range of timescales for which $\varepsilon(t_{E}) > 0$. Following this remark, Kerins (1998) has introduced the concept of observable optical depth defining it as
\begin{equation}
\tau_{oble} = \frac{\pi}{4} \int_{\varepsilon(t_E)>0}{\varepsilon(t_{E}) t_E \frac{d\Gamma}{dt_E} dt_E} \ ,
\label{eq: tauker}
\end{equation}
where $d\Gamma/dt_E$ is the differential rate for the assumed halo model, being the rate $\Gamma$ the number of events per unit time (\cite{MR97}; \cite{J98}). Then, the halo mass fraction $f$ is more correctly estimated as $\tau_{obsd}/\tau_{oble}$, being $\tau_{obsd}$ the directly observed optical depth, i.e. (\cite{K98})
\begin{equation}
\tau_{obsd} = \frac{\pi}{4E} \sum_i t_{E,i} \ .
\label{eq: tauobsd}
\end{equation}
Using these quantities, Kerins has shown that the comparison between observed and obervable optical depths may give only a lower limit on $f$, while no upper limit may be obtained without a knowledge of the full distribution function of the halo model and of the MACHO's mass function (hereafter MF), i.e. the number density of MACHOs with mass in the range $(\mu, \mu + d\mu)$. 
The result obtained by Kerins was proved under the hypothesis that all the lenses have the same mass and assuming a standard cored isothermal sphere for the dark halo mass density. In this paper we generalize the calculation of $\tau_{oble}$ and analyze a general class of homogenous MF. Several studies have been made by many authors to determine the MF function by microlensing data, but they are essentially devoted to the MF of bulge lenses (\cite{Z95}; \cite{HG96}; \cite{Lukas}) and the work is still in progress. Mao \& Paczynski (1996) considered simplified toy models and a power\,-\,law MF and estimated that a reliable determination could be achieved only if we had 100 or more events. Their results, obtained under the assumption that MACHO's spatial distribution and kinematics were known, have been confirmed by Markovic \& Sommer\,-\,Larsen (1997) who have also studied the effect of changing halo model on the estimated average mass of lensing objects. All these studies are based on statistical methods, that is why there is need of a large number of events to reduce the error on parameters determination. In this paper we use a different technique to get useful informations on the lensing objects MF. Since microlensing observable quantities, i.e. number of events, observable optical depth and mean duration, depend on the dark halo model and on the MF, we may express these quantities as functions of some models parameters and of the slope of the MF itself. Then a comparison between theoretical expectations and observed quantities will help us to recover the values of the model's parameters simply imposing that theory and observations are in good agreement. This is what we call the {\it inverse problem in microlensing}. To be meaningful our analysis must take into account also the detection efficiency and that is why our theoretical expectations have to be corrected for this effect too. In the following we will show how this is possible and will get some interesting constraints on the dark halo mass fraction $f$ composed by MACHOs and the slope $\alpha$ of the assumed homogenous power\,-\,law MF for a wide class of spheroidal models. By the way our method is not able to escape the problems connected to the low number of observed events and actually our results are strongly affected by errors which do not allow us to constrain the slope $\alpha$ of the MF in a narrow range. Were the errors be reduced by increasing the number of observed events, our method should be able to narrow the uncertainties on the MACHOs MF. 

There is one possible source of systematic error connected to our analysis. In the previous discussion, we have implicitly assumed that all the observed events are due to MACHOs in the Milky Way dark halo, but it should be considered also the possibility that at least part of the events are due to LMC self lensing. This hypothesis has been suggested in many papers, but the recent analysis by the MACHO group of the spatial distribution of the events across the observed LMC fields has shown that this latter is not compatible with the proposed models of self lensing. As a further test, MACHO has also analyzed the CMD (Color Magnitude Diagram) of the sources of their first eight events to see if they reside in the LMC disk or behind it. Even if the sample is too small to get a definitive result, the hypothesis that all the eight events are due to halo lenses is slightly favoured (\cite {A00c}). However, there is still the possibility that there are no MACHOs at all in the dark halo and that the observed events are due to the LMC dark halo. In fact, the maximum likelihood analysis of the MACHO collaboration has shown that this hypothesis is not completely excluded, even if the needed LMC model should be somewhat extreme. Besides, a recent paper by Alves \& Nelson (2000) has shown that the LMC rotation curve and the data on the kinematics of the carbon stars in LMC are best fitted by a model composed by a flared and warped disk and no dark halo. If this result will be confirmed by future works, it will strenghten our assumption that the observed microlensing events are due to MACHOs in Milky Way dark halo. In the meantime, we are constrained to consider the effects that a possible contamination by self lensing should have on our results. We will discuss how this will affect our results doing some qualitative tests.

In Sect. 2 we introduce and evaluate the microlensing observables, i.e. the number of events, the observable optical depth and the mean duration, taking into account the detection efficiency for models with isotropic maxwellian transverse velocity distribution and homogenous power\,-\,law MF and assuming a wide class of spheroidal non singular isothermal models. The model parameters and the way we fix some of them is detailed in Sect. 3 where we illustrate the models we have chosen to explore. Sect. 4 is devoted to the analysis of the different models with the simple technique of the inverse problem method to get constraints on the slope of the lens MF and the dark halo mass fraction composed by MACHOs. How the self lensing could affect our results is discussed qualitatively in Sect. 5, while a final discussion of the results is presented in the conclusions.

\section{Microlensing observables for spheroidal models with power\,-\,law MF} 

To apply the method we have outlined in the introduction, we have first to calculate the observable quantities in microlensing to which we will compare the observed ones towards LMC. These quantities are the number of events, the observable optical depth and the mean duration. In all this calculation, we will take into account the detection efficiency in order to be sure that the comparison between predicted and observed quantities is meaningful. 

The starting point is the general expression of the differential rate (\cite{DeR91})
\begin{equation}
d\Gamma = 2 D_s r_E v_{\perp} f(v_{\perp}) [\mu s(1-s)]^{1/2}
\frac{dn}{d\mu} d\mu dv_{\perp} ds \ .
\label{eq: dgamma}
\end{equation}
In Eq.(\ref{eq: dgamma}), $f(v_{\perp})$ is the transverse velocity distribution and $dn/d\mu(s,\mu)$ is the halo lens MF; in the reasonable hypothesis of homogeneity, this latter may be factorized as\,:
\begin{equation}
\frac{dn}{d\mu}(s,\mu) = H(s) \times \frac{dn_0}{d\mu}(\mu) =
\frac{\rho(s)}{\rho_{\odot}} \times \frac{dn_0}{d\mu}(\mu)
\label{eq: mffact}
\end{equation} 
being $\rho_{\odot}$ the local mass density of the model and $dn_0/d\mu$ the local MF.  Changing variable from $v_{\perp}$ to $t_E$ and integrating we get
\begin{displaymath}
\frac{d\Gamma}{dt_E} = \frac{8 D_s r_E^3}{t_E^3}
\int_0^1{\frac{\rho(s)}{\rho_{\odot}} [s(1-s)]^{3/2} ds}
\end{displaymath}
\begin{equation}
\ \ \ \ \ \ \ \ \times
\int_{\mu_l}^{\mu_u}{\mu^{3/2} \frac{dn_0}{d\mu} 
\left. f \right|_{v_{\perp} = \frac{2r_E \sqrt{\mu s(1-s)}}{t_E}} d\mu} \ ,
\label{eq: dgammadtgen}
\end{equation}
being $(\mu_l, \mu_u)$ the lower and upper limit for the mass of MACHOs. From Eq.(\ref{eq: dgammadtgen}) one immediately see that to go on further we need to assign the transverse velocity distribution, the local MF and the mass density of the halo model. 

Since we do not consider anisotropy in the velocity space, we may assume the following maxwellian distribution of the transverse velocities
\begin{equation}
f(v_{\perp}) = \frac{2 v_{\perp}}{v_H^2} 
\exp{\left \{ - v_{\perp}^2/v_H^2 \right \}} \ ,
\label{eq: fvperp}
\end{equation}
where $v_{H}$ is the velocity dispersion which we fix as 210 km/s.

With regard to the local MF, it is usual to assume that all the MACHOs have the same mass which means that $dn_0/d\mu$ is a $\delta$\,-\,Dirac centred on the common mass. This is just a first approximation: it is worthwhile to explore different possibilities. As a generalization we consider the case of a homogenous power\,-\,law MF for the MACHOs, i.e. we assume
\begin{equation}
\frac{dn_0}{d\mu}(\mu) = C(\alpha) \mu^{-\alpha} \ ,
\label{eq: mfpowlaw}
\end{equation}
being $C(\alpha)$ a normalization constant fixed such that
\begin{equation}
C(\alpha) \int_{\mu_l}^{\mu_u}{\mu^{1 - \alpha} d\mu}
= \frac{\rho_{\odot}}{M_{\odot}} \ ;
\label{eq: norm}
\end{equation}
this gives :
\begin{equation}
C(\alpha) = \frac{\rho_{\odot}}{M_{\odot}} \times
\left \{
\begin{array}{ll}
\frac{2-\alpha}{\mu_u^{2-\alpha} - \mu_l^{2-\alpha}} & (\alpha \ne 2)\\
\frac{1}{\ln{\mu_u} - \ln{\mu_l}} & (\alpha = 2)\\
\end{array}
\right .
\ .
\label{eq: cnorm}
\end{equation}
 
Concerning the mass density of the dark halo we will restrict our analysis to a class of spheroidal non singular isothermal models whose density distribution is given by
\begin{equation}
\rho(R,z) = \rho_{\odot}^{(s)} \frac{e}{q \sin^{-1}{e}} 
\frac{R_0^2 + R_c^2}{R^2 + z^2/q^2 + R_c^2} \ ,
\label{eq: rho}
\end{equation}  
where $e = \sqrt{1-q}$ with $q$ the halo flattening, $\rho_{\odot}^{(s)}$ is the local mass density\footnote{Note that the $\rho_{\odot}$ previously introduced and $\rho_{\odot}^{(s)}$ are equal only for models with $q = 1$; in general, the link between these two quantities is given by: $\rho_{\odot} = \rho_{\odot}^{(s)} \frac{e}{q \sin^{-1}{e}}$.} for the spherical case, $R_0$ is the galactocentric distance of the Sun and $R_c$ is the core radius. 

Inserting now Eqs.(\ref{eq: fvperp}), (\ref{eq: mfpowlaw}) and (\ref{eq: rho}) into Eq.(\ref{eq: dgammadtgen}) and expressing $(R,z)$ in terms of $(s, l, b)$ (with $(l,b)$ galactic angular coordinates of the target), we finally get
\begin{displaymath}
\frac{d\Gamma}{dt_E} = \frac{2^{(2\alpha -1)} D_s C(\alpha) r_E^{2(\alpha -1)} (R_0^2 + R_c^2)}{v_H^{2(\alpha -2)} t_E^{2(\alpha - 1)}} 
\end{displaymath}
\begin{equation}
\ \ \ \ \ \ \ \ \times
\int_0^1{\frac{[s(1-s)]^{\alpha -1}}{As^2 + Bs + C} \tilde{G}(s,t_E) ds} \ ,
\label{eq: dgammadtfin}
\end{equation}
being
\begin{displaymath}
A = D_s^2 (\cos^2{b} + q^{-2} \sin^2{b}) \ , 
\end{displaymath}
\begin{displaymath}  
B = -2 D_s R_0 \cos{b} \cos{l} \ ,  
\end{displaymath}
\begin{displaymath}
C = R_0^2 + R_c^2 \ ,
\end{displaymath}
and we have defined for sake of shortness
\begin{equation}
\tilde{G}(s,t_E) = G(3-\alpha, \beta(s,t_E) \mu_l) - G(3-\alpha, \beta(s,t_E) \mu_u)
\label{eq: gtilde}
\end{equation}
with
\begin{displaymath}
\beta(s, t_E) = \frac{4 r_E^2 s (1-s)}{v_H^2 t_E^2} \ , \ 
G(a,\zeta) = \int_0^{\zeta}{t^{a-1}\exp{\{-t \}}dt} \ .
\end{displaymath}

Given $\rho(R, z)$ it is quite straightforward to get all the observable quantities we need in the following. The first one is the number of observable events, which is simply given by
\begin{equation}
N_{ev}^{oble} = E \int_{\varepsilon(t_E)>0}
{\varepsilon(t_E) \frac{d\Gamma}{dt_E}dt_E} \ .
\label{eq: nevobledef}
\end{equation}
Introducing the expression given by (\ref{eq: dgammadtfin}) into Eq.(\ref{eq: nevobledef}) one gets
\begin{equation}
N_{ev}^{oble} = K_{ev}(\alpha) I_{ev}(\alpha; \ model) \ ,
\label{eq: nevoble}
\end{equation}
having posed
\begin{equation}
K_{ev}(\alpha) = \frac{2^{(2\alpha -1)} D_s C(\alpha) E r_E^{2(\alpha -1)} (R_0^2 + R_c^2)} {v_H^{2(\alpha -2)}} \ ,
\label{eq: kev}
\end{equation}
\begin{displaymath}
I_{ev}(\alpha; \ model) = \int_{\varepsilon(t_E)>0}{\varepsilon(t_E) t_E^{-2(\alpha -1)} dt_E }
\end{displaymath}
\begin{equation}
\ \ \ \ \ \ \ \ 
\ \ \ \ \ \ \ \ \ \times
\int_0^1{\frac{[s(1-s)]^{\alpha -1}}{As^2+Bs+C} \tilde{G}(s,t_E) ds} \ .
\label{eq: iev}
\end{equation}
Also, from Eq.(\ref{eq: dgammadtfin}) and definition (\ref{eq: tauker}), we get the following expression for the observable optical depth:
\begin{equation}
\tau_{oble} = K_{\tau}(\alpha) I_{\tau}(\alpha; \ model) \ ,
\label{eq: tauoble}
\end{equation}
being now
\begin{equation}
K_{\tau}(\alpha) = \frac{2^{(2\alpha - 3)} \pi D_s C(\alpha) r_E^{2(\alpha -1)}(R_0^2+R_c^2)}{v_H^{2(\alpha -2)}} \ ,
\label{eq: ktau}
\end{equation}
\begin{displaymath}
I_{\tau}(\alpha; \ model) = \int_{\varepsilon(t_E)>0}{\varepsilon(t_E) t_E^{3-2\alpha} dt_E }
\end{displaymath}
\begin{equation}
\ \ \ \ \ \ \ \ 
\ \ \ \ \ \ \ \ \ \times
\int_0^1{\frac{[s(1-s)]^{\alpha -1}}{As^2+Bs+C} \tilde{G}(s,t_E) ds} \ .
\label{eq: itau}
\end{equation}
Now it is not difficult to get the predicted mean duration; in order to take into account the detection efficiency, we define this quantity as
\begin{equation}
< t_E > = \frac{1}{N_{ev}^{oble}} \int_{\varepsilon(t_E)>0}
{t_E dN_{ev}^{oble}} \ .
\label{eq: temediodef}
\end{equation} 
This definition is nothing else but a straightforward generalization of the usual one (\cite{J98}) to which it reduces in the case of perfect detection efficiency, i.e. when $\varepsilon(t_E) = 1 \ \forall \ t_E$. It is not surprising, then, that the following relation holds:
\begin{displaymath}
\tau_{oble} = \frac{\pi}{4E} < t_E > N_{ev}^{oble} \ ,
\end{displaymath}
which is immediately demonstrated. Using such a relation and Eqs.(\ref{eq: nevoble}) and (\ref{eq: tauoble}), we finally get
\begin{equation}
< t_E > = \frac{I_{\tau}}{I_{ev}}(\alpha; \ model) \ .
\label{eq: temedio}
\end{equation} 
In Eqs.(\ref{eq: nevoble}), (\ref{eq: tauoble}) and (\ref{eq: temedio}) we have indicated the unknown parameters which these quantities depend on. These are essentially the slope $\alpha$ of the MF and the parameters needed to specify the characteristics of the mass model. We have decided to not consider the mass limits as unknown parameters, although they are not very well constrained. In fact, in the following analysis we retain as fixed these two quantities and investigate the dependence on the remaining unknowns. 

\section{Exploring the models parameters space}

\begin{table}
\caption{Models main characteristics. The parameter $\rho_{\odot}^{(s)}$ is in units of $10^{-3} \ M_{\odot} \ pc^{-3}$; $M_{50}$ is the total mass inside 50 kpc in units of $10^{11} \ M_{\odot}$ and $v_c(\infty)$ the asymptotic rotation velocity in km/s.}
\begin{center}
\begin{tabular}{|c|c|c|c|} 
\hline q & $\rho_{\odot}^{(s)}$ & $M_{50}$ & $v_c(\infty)$ \\
\hline
0.3 & 6.7 & 3.90 & 197 \\
0.5 & 7.2 & 4.02 & 201 \\
0.8 & 8.1 & 4.21 & 208 \\
1.0 & 8.7 & 4.32 & 212 \\
\hline
\end{tabular}
\end{center}
\end{table} 

The integrals in Eqs.(\ref{eq: iev}) and (\ref{eq: itau}) may be evaluated only numerically, but to do this we need to fix all the parameters entering the mass density in Eq.(\ref{eq: rho}) (i.e. $q$, $\rho_{\odot}^{(s)}$, $R_0$, $R_c$), the lower and upper mass limits $(\mu_l, \mu_u)$ and also the slope $\alpha$ of the MF which is just the parameter we want to determine. 

We have fixed $R_0 = 8.0$\,kpc and $R_c = 5.6$\,kpc (as usual in literature). Next we have to choose the halo flattening $q$, but the constrains on it are really very poor. Different kinds of analysis with different techniques give very different results (for a review see, e.g., \cite{R96} and \cite{S99}) and there is no general agreement on which value is the best one to use. Since there is such a large uncertainty, we have decided to consider four values of the halo flattening, repeating our analyses for models with $q = 0.3, 0.5, 0.8, 1.0$ in order to test if this parameter has some effect on the results. The last parameter to fix is the local mass density for the spherical case $\rho_{\odot}^{(s)}$. This is fixed such that the local rotation velocity is equal to the observed value of $v_c(R_0) = 200$\,km/s. To this aim we have also considered the contributions of the bulge and disk. We modelled these components as in M\'era {\it et al.} (1998): the bulge is treated as point like with total mass $1.23 \times 10^{10} \ M_{\odot}$ and the disk as a double exponential with scale length $R_d = 3.5$\,kpc and local surface mass density $\Sigma_{\odot} = 52 \ M_{\odot} pc^{-2}$. In Table 1 we give the values of the models parameters together with the total mass inside 50 kpc and the asymptotic rotation velocity. We would like to note that the predicted values of these latter quantities are in good agreement within the errors with the recent measurements (see, e.g., \cite{WE99}) for all the considered models.

\begin{table*}
\caption{Values of the coefficients entering the numerical approximations given in Eqs.(\ref{eq: ievnum}) and (\ref{eq: itaunum}) for models A1 (first half) and A2 (second half).}
\begin{center}
\begin{tabular}{|c|c|c|c|c|c|c|c|c|} 
\hline q & $a_{ev}$ & $b_{ev}$ & $c_{ev}$ & $d_{ev}$ & $a_{\tau}$ & $b_{\tau}$ & $c_{\tau}$ & $d_{\tau}$ \\
\hline \hline
0.3 & 0.08 & 0.22 & -8.83 & 3.42 & 0.13 & -0.09 & -8.67 & 7.00 \\
0.5 & 0.08 & 0.23 & -8.84 & 3.94 & 0.13 & -0.09 & -8.67 & 7.57 \\
0.8 & 0.08 & 0.24 & -8.84 & 4.26 & 0.13 & -0.09 & -8.67 & 7.92 \\
1.0 & 0.08 & 0.24 & -8.84 & 4.36 & 0.13 & -0.09 & -8.67 & 8.02 \\
\hline
0.3 & 0.17 & 0.35 & -11.33 & 7.48 & 0.29 & -0.47 & -10.67 & 11.92 \\
0.5 & 0.17 & 0.38 & -11.35 & 7.98 & 0.29 & -0.47 & -10.67 & 12.47 \\
0.8 & 0.17 & 0.39 & -11.36 & 8.29 & 0.29 & -0.46 & -10.68 & 12.81 \\
1.0 & 0.16 & 0.40 & -11.37 & 8.38 & 0.29 & -0.46 & -10.68 & 12.91 \\
\hline
\end{tabular}
\end{center}
\end{table*}

\begin{table*}
\centering
\caption{The same as in Table 2 but for models B1 (first half) and B2 (second half).}
\begin{tabular}{|c|c|c|c|c|c|c|c|c|} 
\hline q & $a_{ev}$ & $b_{ev}$ & $c_{ev}$ & $d_{ev}$ & $a_{\tau}$ & $b_{\tau}$ & $c_{\tau}$ & $d_{\tau}$ \\
\hline \hline
0.3 & 0.008 & 0.16 & -8.75 & 3.38 & 0.02 & 0.10 & -8.81 & 7.02 \\
0.5 & 0.008 & 0.16 & -8.75 & 3.89 & 0.02 & 0.10 & -8.81 & 7.58 \\
0.8 & 0.009 & 0.16 & -8.74 & 4.21 & 0.02 & 0.10 & -8.81 & 7.93 \\
1.0 & 0.008 & 0.16 & -8.75 & 4.31 & 0.02 & 0.10 & -8.81 & 8.04 \\
\hline
0.3 & 0.05 & 0.42 & -11.23 & 7.44 & 0.12 & -0.01 & -11.02 & 11.98 \\
0.5 & 0.05 & 0.43 & -11.23 & 7.93 & 0.12 & 0.004 & -11.02 & 12.53 \\
0.8 & 0.05 & 0.43 & -11.24 & 8.24 & 0.12 & 0.007 & -11.02 & 12.87 \\
1.0 & 0.05 & 0.43 & -11.24 & 8.34 & 0.12 & 0.008 & -11.02 & 12.97 \\
\hline
\end{tabular}
\end{table*}

In order to obtain the observable in term of $\alpha$ we have numerically integrated Eqs.(\ref{eq: iev}) and (\ref{eq: itau}) for many values of $\alpha$ and then interpolated the results to get $I_{ev}$ and $I_{\tau}$ as functions of $\alpha$ itself. We have then repeated this procedure changing the values for the halo flattening $q$ and the mass limits in order to investigate a wide class of halo models, each one labelled with a code given as follows. We named A1, A2 models with $\mu_l = 0.001$ and $\mu_u = 0.1$ and $\mu_u = 1.0$ respectively, and  B1, B2 models with $\mu_l = 0.01$ and $\mu_u = 0.1$ and $\mu_u = 1.0$ respectively. Then we add a letter to indicate the halo flattening with the following conventions: $a \rightarrow q = 0.3$, $b \rightarrow q = 0.5$, $c \rightarrow q = 0.8$, $d \rightarrow q = 1.0$. So, e.g., the model labelled A2c has: $\mu_l = 0.001$, $\mu_u = 1.0$, $q = 0.8$. Thus, we consider sixteen different models in the same class.

Before going on, we would like to discuss how we have chosen the mass limits $(\mu_l, \mu_u)$. Concerning the upper limit, it has been known for a long time now (\cite{GH83}) that hydrogen\,-\,burning stars cannot provide the majority of the halo dark matter. Numerous recent studies (\cite{B94}; \cite{Hu94}; \cite{GF96}) put an upper limit of at most 4\% on the dark halo density contribution of hydrogen\,-\,burning stars. For an old, metal\,-\,poor population this means that stars with mass between 0.1\,$M_{\odot}$ and 0.8\,$M_{\odot}$ give no significant contribution to the dark matter in the galactic halo. These evidences suggest to fix $\mu_u = 0.1$, but we have decided to consider also models with $\mu_u = 1.0$ for the reasons we are going to explain. On one hand, MACHO (\cite{A00}) and EROS (\cite{R97}) results indicate that the most likely MACHO' s mass is $0.5 \ M_{\odot}$. On the other hand,  Kerins (1997) has shown that MACHOs may reside in a population of dim halo globular clusters comprising mostly or entirely low\,-\,mass stars just above the hydrogen\,-\,burning limit. For the case of the standard halo model, this scenario is consistent not only with MACHO observations, but also with cluster dynamical constraints and number counts limits imposed by twenty HST fields. Further suggestions of the possible existence of MACHOs with mass $\sim 0.5 \ M_{\odot}$ come from the study of the double quasars variability (\cite{KdB99}). All these studies have led us to consider also models with $\mu_u = 1.0$. 

Fixing the lower limit $\mu_l$ is not an easy task too. De Rujula et al. (1991) have shown that a lower limit for the mass of MACHOs is $10^{-7} \ M_{\odot}$, but this does not imply that objects with a mass $\sim 10^{-7} \ M_{\odot}$ really exist. Actually, MACHO and EROS search for short duration events pose strong constraints on their contribution to the halo mass budget (\cite{A96}; \cite{A98}; \cite{R98}). Following these works, we have chosen two values for $\mu_l$ given by 0.001 and 0.01 respectively. In all our analysis we are assuming that the MF is the same in the mass range $(\mu_l, \mu_u) \ M_{\odot}$, i.e. that the slope $\alpha$ does not change in this range, which seems quite reasonable as a first approximation.

To integrate Eqs.(\ref{eq: iev}) and (\ref{eq: itau}) we need the detection efficiency of the MACHO collaboration for their monitoring campaign towards LMC since in our analysis we will use their results of the first 5.7 years of observations. This function has been carefully evaluated by the MACHO group itself (\cite{epsMACHO}), but they give no analytical formula for it. That is why we have built up an approximated expression of $\varepsilon(t_E)$ interpolating the data taken from Fig. 5 of Alcock {\it et al.} (2000a), obtaining
\begin{equation}
\varepsilon(t_E) =
\left \{
\begin{array}{ll}
0.108 \exp{\{-0.02 (\log{t_E})^{3.8}\}} (\log{t_E})^{2.4} & \\
0.41 \exp{\{-14.7 (\log{t_E} - 2.56)^2\}} & \\
\end{array}
\right .
\ ,
\label{eq: epst}
\end{equation}
being $\log$ the decimal logarithm; the first expression holds for $2 \ d \le t_E \le 300 \ d$ and the second for $300 \ d < t_E \le 900 \ d$. Eq.(\ref{eq: epst}) differs from the measured $\varepsilon(t_E)$ less than 10\% in the range examined, the error being larger for events lasting more than 900 days. This is not a problem since the thirteen observed events which we use in our analysis last approximately from 34 to 102 days, so we are confident that no serious systematic error is induced by our approximation for $\varepsilon(t_E)$. For the same reason also the discontinuity in $t_E = 300$\,d has no effect on our analysis. 

\begin{figure}[ht]
\sidecaption
\resizebox{7.5cm}{!}{\includegraphics{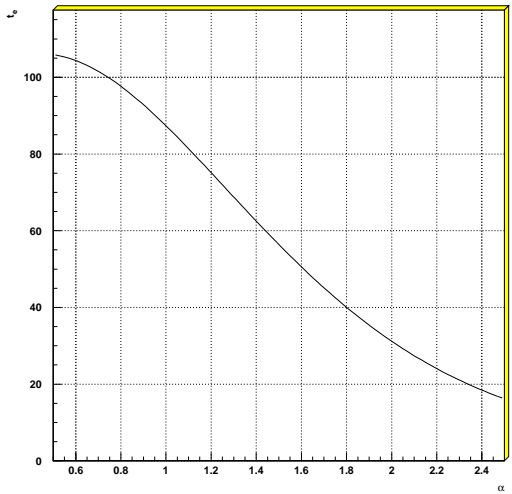}}
\hfill
\caption{$< t_E >(\alpha)$ for the model named A2c, i.e. with $\mu_l = 0.001$, $\mu_u = 1.0$, $q = 0.8$. Similar plots are obtained for the other models.}
\end{figure}

Now we have all we need to estimate the functions defined in Eqs.(\ref{eq: iev}) and (\ref{eq: itau}). Without entering in details, the numerical integration and the following interpolation of the results have shown us that it is possible to write
\begin{equation}
I_{ev}(\alpha) = 
\exp{\{a_{ev} \alpha^3 + b_{ev} \alpha^2 + c_{ev} \alpha + d_{ev}\}}
\ ,
\label{eq: ievnum}
\end{equation}
\begin{equation}
I_{\tau}(\alpha) = 
\exp{\{a_{\tau} \alpha^3 + b_{\tau} \alpha^2 + c_{\tau} \alpha + d_{\tau}\}}
\ ,
\label{eq: itaunum}
\end{equation}
where the values of the coefficients $(a_{ev}, b_{ev}, c_{ev}, d_{ev})$ and $(a_{\tau}, b_{\tau}, c_{\tau}, d_{\tau})$ depend on the model considered. We report them in Tables 2 and 3. These approximations work quite well, the error being always less than 10\,\% and in many cases also less than 5\,\%. As an example, in Fig. 1 we plot the ratio $I_{\tau}/I_{ev} = < t_E >(\alpha)$ for model A2c.

\section{The inverse problem to estimate $(\alpha, f)$}

We show now how it is possible to use the simple technique of the inverse problem to estimate the slope $\alpha$ of the lens MF and the dark halo mass fraction $f$ composed by MACHOs for each one of the models labelled by the codes previously explained. We consider data coming from the first 5.7 years of observations towards LMC by the MACHO collaboration (\cite{A00}), limiting ourselves to the thirteen events selected according to the so\,-\,called {\it selection criteria A}. These are high S/N events and are spatially distributed in a way which is consistent with the hypothesis that they are due to lenses belonging to our halo and not to LMC self lensing; we will discuss of this problem later on. For this set of events we have
\begin{displaymath}
N_{ev}^{obsd} = 13 \ , \ \ 
\tau_{obsd} = 3.5 \times 10^{-8} \ , \ \ < t_E >_{obsd} = 76.2 \ d \ ,
\end{displaymath}
where $< t_E >_{obsd}$ is simply the average value  of the duration of the observed events. As a test of the correctness of this estimate we may note that $(\pi/4E) < t_E >_{obsd}$ (with $E = 6.12 \times 10^{7}$ star\,-\,years) is exactly equal to $\tau_{obsd}/N_{ev}^{obsd}$, as it has to be for the reasons we are going to explain later. We have then to estimate the uncertainties on the observed quantities. First, we consider the directly observed optical depth, and we simply use a method as similar as possible to the one proposed by the MACHO group itself for a very conservative estimate of the error on $\tau_{meas}$ (\cite{A97a}): we divide the observed events according to their duration $t_E$ in bin of 10 days; in such a way $\tau_{obsd}$ is more or less the same for events in the same bin and the errors are approximately poissonian. For each bin we estimate $N_{low} = N_{ev}^{obsd}(bin) - \sqrt{N_{ev}^{obsd}(bin)}$ and $N_{up} = N_{ev}^{obsd}(bin) + \sqrt{N_{ev}^{obsd}(bin)}$ and define $\tau_{1}^{min}$ and $\tau_{1}^{max}$ as the minimum and maximum value of $\tau_{1}$ for the events in that bin (being $\tau_1 = \pi t_E / 4 E$). Then we estimate
\begin{displaymath}
\tau_{obsd}^{low} = \sum_{bin} N_{low} \tau_{1}^{min} \ , \ \ \ \ 
\tau_{obsd}^{max} = \sum_{bin} N_{up} \tau_{1}^{max} \ .
\end{displaymath}
At the end we find
\begin{displaymath}
\tau_{obsd} = (3.5 \pm 2.5) \times 10^{-8} .
\end{displaymath}
The error on the optical depth turns out to be so large ($\sim 70 \%$) because of the limited number of events. Let us turn now to the error on the number of observed events. This simply comes from the low statistics and may be assumed to be poissonian, i.e. $\delta N_{ev}^{obsd}/ N_{ev}^{obsd} = 1/\sqrt{N_{ev}^{obsd}} \simeq 28\%$. Finally, the error on $< t_E >$ is obtained by propagating the error on $N_{ev}^{obsd}$ which gives us 
\begin{displaymath}
\frac{\delta < t_E >_{obsd}}{< t_E >_{obsd}} = 
\frac{\delta N_{ev}^{obsd}}{N_{ev}^{obsd}} \ .
\end{displaymath}

\begin{figure*}[ht]
\centering
\resizebox{13cm}{!}{\includegraphics{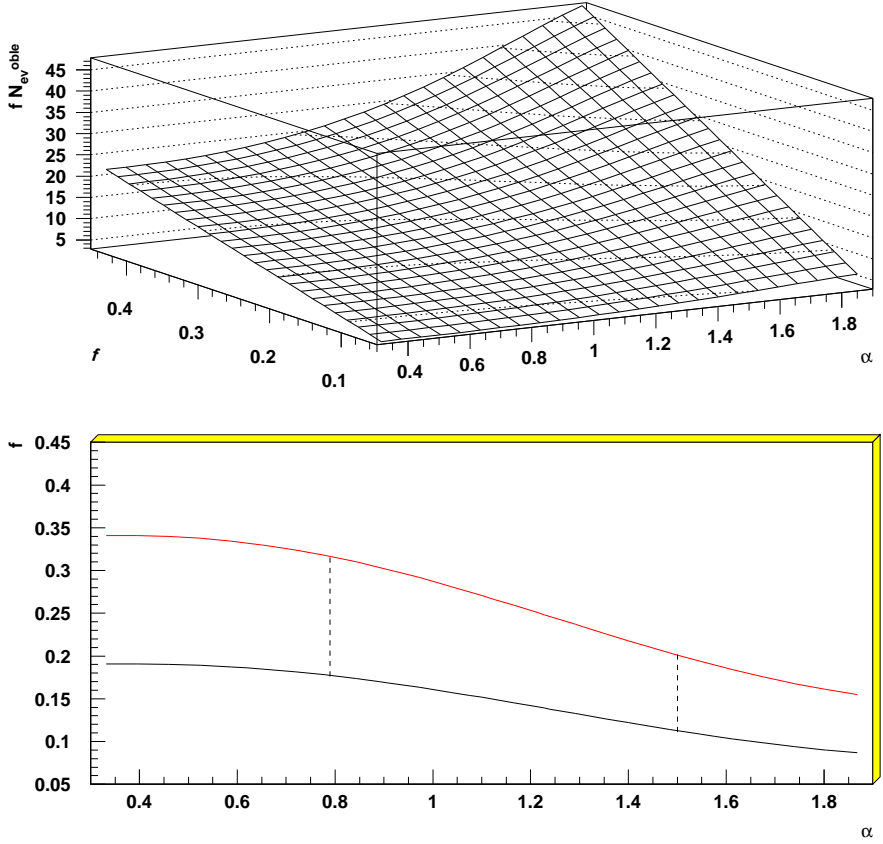}}
\hfill
\caption{In the upper panel we plot $f \times N_{ev}^{oble}$ as a function of $\alpha$ and $f$; in the lower one, we determine the region $\cal{R}$ of the parameter space $(\alpha, f)$ consistent with microlensing observations towards LMC: the upper line in the second panel is the level curve $f \times N_{ev}^{oble} = 16.61$, the lower one for  $f \times N_{ev}^{oble} = 9.39$, whilst the two dashed lines have been drawn intersecting the level curves with the vertical lines $\alpha = 0.79$ and $\alpha = 1.50$. The plot is for model A2c; similar plots are obtained for the other models.}
\end{figure*}

The observed quantities may be now compared to the theoretical ones evaluated in the previous section, which also take into account the detection efficiency in order to make the comparison meaningful. To do this we must first remember that the theoretical quantities have been calculated under the hypothesis that the dark halo is totally made out by MACHOs, i.e. with $f = 1$. Actually, the exact value of $f$ is not well known: from the more recent observational constraints it is quite unlikely that $f = 1$. However to take into account of $f$ is not very difficult: we have simply to multiply by $f$ the expression of the differential rate $d\Gamma/dt_E$  and consequently the ones obtained for $N_{ev}^{oble}$ and $\tau_{oble}$. Note that $< t_E >$ is independent on the value of $f$.
 Then we have the following relations between observable and observed quantities:
\begin{equation}
\left \{
\begin{array}{ll}
N_{ev}^{oble} = N_{ev}^{obsd}/f \\
\tau_{oble} = \tau_{obsd}/f \\
\end{array}
\right .
\ .
\label{eq: system}
\end{equation}
Dividing these two equations and using the relation $\tau_{oble}/N_{ev}^{oble} = (\pi/4E)< t_E >$, one gets: $\tau_{obsd}/N_{ev}^{obsd} = (\pi/4E)< t_E >_{obsd}$, which may be used to test the correctness of our previous estimate of $< t_E >_{obsd}$. For each model, we may solve the system (\ref{eq: system}) in the unknowns $\alpha$ and $f$ and estimate them toghether with the errors connected to our analysis. Solving Eqs.(\ref{eq: system}) is not really useful since the very high error on $\tau_{obsd}$ leads us to get no constraints at all on the parameters $(\alpha, f)$. We may then use a third relation we have at our disposal, given by
\begin{equation}
< t_E >(\alpha) =  < t_E >_{obsd} \ ,
\label{eq: tecompare}
\end{equation}
which is quite easy to solve numerically to get an estimate of the slope $\alpha$ of the MF.  Solving Eq.(\ref{eq: tecompare}), taking into account also the uncertainties, will give us in general more than one solution compatible with the microlensing data. A further selection can be done imposing that $\alpha$ must be in the range $(0.0, 5.0)$ since values outside this range are reasonably quite unlikely\footnote{Negative values of $\alpha$ means that the MF is increasing which has never been observed, whilst values greater than 5.0 are quite unusual.}. In this way, for each model, we have estimated a range $(\alpha_l, \alpha_u)$ for the slope of the MF simply requiring that
\begin{displaymath}
(1 - \delta t_E / t_E ) < t_E >_{obsd} \le < t_E >(\alpha) \le 
\end{displaymath}
\begin{displaymath}
(1 + \delta t_E / t_E ) < t_E >_{obsd} \ \iff 
\end{displaymath}
\begin{equation}
54.86 \ d \le < t_E >(\alpha) \le 97.54 \ d \ 
\label{eq: condte}
\end{equation}
being $\delta t_E/ t_E$ the fractional error on $< t_E >_{obsd}$ which we have previously estimated to be of order of 28\,\%. These values are summarised in Table 4, where we report also the value of $\alpha_0$ corresponding to the average value $< t_E >(\alpha_0) = 76.20$\,d.

\begin{table*}
\caption{Estimates of the slope $\alpha$ and the dark halo mass fraction $f$ composed by MACHOs for different models labelled as explained before. $M_{50}^{bar}$ is the mass in MACHOs inside 50 kpc (with $M_{50,(0)}^{bar}$ the value corresponding to $\alpha_0$ and $f_0$) measured in units of $10^{10} \ M_{\odot}$. Note that the value of $\alpha_l$ for model B2a differs considerably from the others since the maximum value of $< t_E >(\alpha)$ is only 96.2 d, which is lower than the upper limit on the $< t_E >_{obsd}$ given in Eq.(\ref{eq: condte}).}
\begin{center}
\begin{tabular}{|c|c|c|c|c|} 
\hline Code & $(\alpha_l, \alpha_u)$ & $(f_{min}, f_{max})$ & $M_{50}^{bar}$ & $(\alpha_0, f_0, M_{50,(0)}^{bar})$ \\
\hline \hline
A2a & (0.50, 1.48) & (0.12, 0.33) & 4.68 $\div$ 12.87 & (1.10, 0.21, 8.19) \\
A2b & (0.73, 1.50) & (0.10, 0.30) & 4.02 $\div$ 12.06 & (1.15, 0.19, 7.63) \\
A2c & (0.80, 1.50) & (0.11, 0.32) & 4.63 $\div$ 13.47 & (1.17, 0.20, 8.42) \\
A2d & (0.84, 1.56) & (0.12, 0.35) & 5.18 $\div$ 15.12 & (1.21, 0.22, 9.50) \\
\hline
B2a & (0.27, 1.69) & (0.11, 0.33) & 4.29 $\div$ 12.87 & (1.12, 0.24, 9.36) \\
B2b & (0.63, 1.80) & (0.10, 0.30) & 4.02 $\div$ 12.06 & (1.25, 0.18, 7.24) \\
B2c & (0.77, 1.89) & (0.10, 0.32) & 4.21 $\div$ 13.47 & (1.34, 0.19, 8.00) \\
B2d & (0.77, 1.90) & (0.11, 0.34) & 4.75 $\div$ 14.79 & (1.34, 0.21, 9.07) \\
\hline
\end{tabular}
\end{center}
\end{table*} 

Having estimated $\alpha$ and being $N_{ev}^{obsd}$ known, we may now get a constraint also on $f$: on the plot $f N_{ev}^{oble}$ as a function of $\alpha$ and $f$ we make the contour levels for $f N_{ev}^{oble} = N_{ev}^{obsd} - 28\,\% = 9.39$ and $f N_{ev}^{oble} = N_{ev}^{obsd} + 28\,\% = 16.61$. On this graph one has to add also the vertical lines corresponding to $\alpha = \alpha_l$ and $\alpha = \alpha_u$\ (see Fig. 2). The region $\cal{R}$ of the parameter space $(\alpha, f)$ delimited by the two level curves and these two vertical lines is that which is consistent with the constraints on $\alpha$ and the ones coming from microlensing observations towards LMC, i.e. the number of observed events and their mean duration. From Fig. 2 one sees that it is possible to define $f_{min}$ and $f_{max}$ as the minimum and maximum value of $f$ in the region $\cal{R}$ and a value $f_0$ such that
\begin{displaymath}
\left \{
\begin{array}{ll}
f_0 N_{ev}^{oble}(\alpha_0) = 13 \\
\\
< t_E >(\alpha_0) = 76.20 \ d \\
\end{array}
\right .
\ .
\end{displaymath}
One has then that for each value of $f \in (f_{min}, f_{max})$ there exists a value of $\alpha \in (\alpha_l, \alpha_u)$ such that
\begin{equation}
\left \{
\begin{array}{ll}
9.39 \le f N_{ev}^{oble}(\alpha) \le 16.61 \\
\\
54.86 \ d \le < t_E >(\alpha) \le 97.54 \ d \\
\end{array}
\right .
\ .
\label{eq: constraints}
\end{equation}

The values of $(f_{min}, f_{max})$ and $f_0$ are summarised in Table 4 for the different models considered\footnote{A complete list of the solutions for each model, comprising also the rejected ones, and further details on the numerical procedures are available on request to the authors.}. We also give in the same table the range for the mass in MACHOs inside 50 kpc (indicated as $M_{50}^{bar}$) which is easily estimated as $f \times M_{50}$ (with $M_{50}$ given in Table 1) since it is this quantity which is most strongly constrained by microlensing observations.

\section{The self lensing problem}

We discuss now the problem of the LMC self lensing and show how our results depend quite weakly on the systematics connected with it LMC. As discussed in the introduction, our analysis implicitly assumed that all the observed events are due to lenses belonging to our dark halo. Even if there are different evidencies against a dominant contribution of self lensing, it is still possible that two or three of the 13 events considered are due to LMC self lensing. This means that we have to repeat our analysis excluding from the sample those events which are not caused by MACHOs. Unfortunately, it is not really possible to establish precisely what are these events, due to the well known degeneracy in the lens parameters which do not allow to determine its distance with respect to the observer. However, we may qualitatevely correct our results excluding form the sample some events chosen according to which are more likely to be due to self lensing. To this aim we have used the results on the self lensing events timescale distribution obtained by Gyuk, Dalal and Griest (2000). Their analysis shows that the most likely duration of self lensing events is $\sim 100$\,d; so we repeated our analysis for different choices of the excluded events. As an example we discuss here the results obtained for the model A2c $(\mu_l = 0.001, \mu_u = 1.0, q = 0.8)$; similar results are obtained for the other models. As a first test, we have excluded from the sample the events labelled 7, 13 and 14 in Table 7 of Alcock et al. (2000a) which are the longest ones. We have then $< t_E >_{obsd} = 68.78 \ d$ and $N_{ev}^{obsd} = 10$; the uncertainties on these quantities increase to $\sim 32 \%$ both on $< t_E >_{obsd}$ and on $N_{ev}^{obsd}$. We get
\begin{displaymath}
\alpha = 0.94 \div 1.67 \  ; \ f = 0.07 \div 0.24 \ ; \ 
(\alpha_0 , f_0) = (1.30, 0.14) .  
\end{displaymath}
Even if there is a little trend towards larger values of $\alpha$ and smaller values of $f$, these results are quite consistent with the ones given in Table 4. The central value for the slope $\alpha$ is slightly larger and $f_0$ slightly lower but the discrepancies are not significative. We have repeated the same analysis in other two cases: excluding events 5, 7 and 13 and exlcuding events 13, 21, 25 respectively. The results we get are consistent with each other and with the results for model A2c in Table 4. Finally we have also tried to exclude four events (events number 7, 13, 14 and 21) from the sample; the uncertainties are of course larger ($\sim 33\%$ both on $< t_E >_{obsd}$ and on $N_{ev}^{obsd}$) while the mean duration is 66.07 d for 9 events. We obtain
\begin{displaymath}
\alpha = 0.99 \div 1.72 \ ; \ f = 0.06 \div 0.21 \ ; \ 
(\alpha_0 , f_0) = (1.34, 0.12) .  
\end{displaymath}
These results are still consistent with those obtained till now for model A2c. These qualitative tests are encouraging since they suggest that our analysis is not seriously affected by the systematics connected to the self lensing whose main effect seems to be to lower the statistics.

\section{Discussion of results and conclusions}

In this paper we have analysed a wide class of spheroidal non singular isothermal dark halo models with a homogenous power\,-\,law MF by changing the halo flattening $q$ and the lower and upper MACHO mass limits $(\mu_l, \mu_u)$. Using the simple technique of the inverse problem method we have obtained useful constraints on the slope $\alpha$ of the MF and the dark halo mass fraction $f$ made out by MACHOs. The results obtained are summarized in Table 4, where we report the estimated ranges for the slope $\alpha$ of the lens MF, for the the dark halo mass fraction $f$ made out by MACHOs and for the mass in MACHOs inside 50 kpc. In the same Table we also report the values $(\alpha_0, f_0, M_{50,(0)}^{bar})$ (the latter being the value of $M_{50}^{bar}$ obtained with $f = f_0$) which may be considered as a sort of {\it best fit} for each model, since for these values of the parameters the model is consistent with microlensing constraints independently on the estimates of the errors. Having in mind that for each value of $f \in (f_{min}, f_{max})$ there exists only one value of $\alpha \in (\alpha_l, \alpha_u)$ such that the model with these values of $(\alpha, f)$ satisfies the constraints (\ref{eq: constraints}), we may draw some interesting considerations from Table 4. 

\begin{enumerate}

\item{The first striking feature in Table 4 is that there are no models with $\mu_u = 0.1$ (models A1 and B1 in our notation). This happens because the constraint (\ref{eq: condte}) on $< t_E >(\alpha)$ for these models gives values of $\alpha$ outside the range $(0.0, 5.0)$. Thus we may conclude that non singular isothermal spheroidal models filled with MACHOs of mass less than $0.1 \ M_{\odot}$ are not consistent with microlensing observations towards LMC. Thus there exist MACHOs with mass greater than the hydrogen burning limit ($\sim 0.1 \ M_{\odot}$); such interpretation is consistent with MACHO and EROS result on the most likely MACHO mass ($\sim 0.5 \ M_{\odot}$) and will have important implications on the question of their nature. Actually, if MACHOs weigth more than $0.1 \ M_{\odot}$, then they should be unusual baryonic objects (e.g. very old white dwarfs) or non baryonic at all (\cite{Sazhin}; \cite{GZS98}). However, data are too few to conclude that this interpretation is correct. We have examined spheroidal non singular isothermal density models, but it is still possible that the halo mass density has a different radial profile with a non flat rotation curve (see, e.g., \cite{WE99}) which deserve some attention.}

\item{The estimated ranges for $\alpha$ do overlap both for models with the same halo flattening $q$ but different mass limits $(\mu_l, \mu_u)$ and for models with the same mass limits but different halo flattening. This is encouraging since it means that our estimates do not depend on the halo flattening which is poorly constrained. Considering the whole set of models in Table 4, an estimate of $\alpha$ can be given in the range $0.27 \div 1.90$, which practically is the widest range of variability. Unfortunately it is not a narrow range because of the high errors connected to the low statistics. 

It is not very easy to compare this estimate with other results, since the present state of the studies on the dark halo MF is still in progress (\cite{MCS98}). Collecting together results on the nearby halo LF, based on geometric parallax determinations of high velocity faint stars in the solar neighbourhood, and those about the spheroid MF, Chabrier \& M\'era estimate that the halo MF is a power\,-\,law with $\alpha = 1.7 \pm 0.2$ (\cite{CM97}). We can see from Table 4 that the range for $\alpha$ obtained for the B2 models, that is for $\mu_l= 0.01$ and $\mu_u= 1.0$, are consistent with the value obtained by Chabrier \& M\'era (1997). This means to put a constrain on the low mass objects of the dark halo. Another important indication on the halo MF comes from studies of the stellar populations of the globular clusters. The MFs of different clusters observed with the HST have been determined recently by Chabrier \& M\'era (1997) who found that they are weakly dependent on the metallicity and may be modelled as power\,-\,law with slope $\alpha \simeq 0.5 \div 1.5$ till 0.1\,$M_{\odot}$. With this range for $\alpha$ also the A2 models are consistent. Note that we have also consistency with MACHO results.}

\item{The estimated ranges for $f$ are consistent with the recent estimate obtained by MACHO (\cite{A00}) using a maximum likelihood method and a different set of halo models. The {\em best fit} values $f_0$ are constrained in a very narrow range: from 0.19 to 0.24; the analysis of Table 4 shows that the total range for $f$ is quite narrow as well: from 0.10 to 0.35. This shows that it is not possible to increase the dark halo mass fraction flattening the dark halo. The same agreement with MACHO results is found with respect to the baryonic mass inside 50 kpc ($M_{50}^{bar}$ in Table 4. Our estimates range from 4.02 to $15.12 \times 10^{10} \ M_{\odot}$ whilst MACHO estimates range from 6 to $13 \times 10^{10} \ M_{\odot}$. Also the {\em best fit} values are in good agreement. These are encouraging results and lead us to be quite confident in our analysis. Once again, one should also consider different radial profiles for the mass density before drawing some definitive conclusion.}

\end{enumerate}

Although not definitely conclusive, we obtained constrains on the dark halo parameters on the basis of microlensing observations for a wide class of spheroidal non singular isothermal models. We performed our analysis on the basis of the microlensing data only in one direction, the LMC, and in fact it has been not possible to determine uniquely the properties of the models we considered. In order to investigate the shape of the halo and put more precise constraints on the relative parameters, it will be necessary to have microlensing data relative to observations towards other directions, for example towards globular clusters; they may be used as sources (\cite{GH97}) or as sites of lenses when observing towards SMC (\cite{JSW98}). There are also other possible targets such as spiral arms (\cite{Lukas}) which could be investigated. It would be also interesting to compare microlensing results on galactic models parameters with the ones obtained from the rotation curves of other spiral galaxies similar to our Milky Way, in particular those deduced from the {\it universal rotation curve} proposed by Persic, Salucci \& Stel (1996) upon their analysis of a large homogenous sample of spiral galaxies. All that constitutes material for further work.

\begin{acknowledgements}
It is a great pleasure to thank the SLOTT group of Napoli (G. Covone, E. Piedipalumbo, C. Rubano, P. Scudellaro and M. Sereno) and Salerno (V. Bozza, S. Calchi Novati, S. Capozziello, G. Iovane, G. Lambiase, V. Re and G. Scarpetta) and Ph. Jetzer and L. Grenacher for the discussions we had on the manuscript. We also thank the anonymous referee for the usefull comments which have helped to improve the paper.
\end{acknowledgements}


\begin{thebibliography}{}

\bibitem[Afonso et al. 1999]{Af98}
Afonso, C. et al. 1999,A\&A, 344, L63 

\bibitem[Alcock et al. 1996]{A96}
Alcock, Ch. et al. 1996, ApJ, 471, 774

\bibitem[Alcock et al. 1997a]{A97a}
Alcock, Ch. et al. 1997a, ApJ, 486, 697

\bibitem[Alcock et al. 1997b]{A97b}
Alcock, Ch. et al. 1997b, ApJ, 479, 119

\bibitem[Alcock et al. 1997c]{A97c}
Alcock, Ch. et al. 1997c, ApJ, 491, L11

\bibitem[Alcock et al. 1998]{A98}
Alcock, Ch. et al. 1998, ApJ, 499, L9

\bibitem[Alcock et al. 2000a]{A00}
Alcock, Ch. et al. 2000a, ApJ, 542, 281

\bibitem[Alcock et al. 2000b]{epsMACHO}
Alcock, Ch. et al. 2000b, astro-ph/0003392

\bibitem[Alcock et al. 2000b]{A00c}
Alcock, Ch. et al. 2000c, astro-ph/0008282

\bibitem[Alves \& Nelson 2000]{AN00}
Alves, D.R., Nelson, C.A. 2000, ApJ, 542, 789

\bibitem[Ansari et al. 1997]{Ans97}
Ansari, R. et al. 1997, A\&A, 324, 843

\bibitem[Bahcall et al. 1994]{B94}
Bahcall, J.N., Flynn, C., Gould, A., Kirhakos, S. 1994, ApJ, 435, L51

\bibitem[Chabrier \& M\'era 1997]{CM97}
Chabrier, G., M\'era, D. 1997, A\&A, 328, 83

\bibitem[De Rujula et al. 1991]{DeR91}
De Rujula, A., Jetzer, Ph., Mass\`o, E. 1991, MNRAS, 250, 348

\bibitem[Flynn et al. 1996]{FGB96}
Flynn, C., Gould, A., Bahcall, J. N. 1996, ApJ, 466, L55

\bibitem[Gates et al. 1995]{GGT95}
Gates, E. I., Gyuk, G., Turner, M. S. 1995, Phys. Rev. Lett., 74, 3274

\bibitem[Gilmore \& Hewett 1983]{GH83}
Gilmore, G., Hewett, P. 1983, Nature, 306, 669

\bibitem[Gould 1996]{G96}
Gould, A. 1996, PASP, 108, 465

\bibitem[Graff \& Freese 1996]{GF96}
Graff, D. S., Freese, K. 1996, ApJ, 456, L49 

\bibitem[Griest 1991]{Gr91}
Griest, K. 1991, ApJ, 366, 412

\bibitem[Grenacher et al. 1999]{Lukas}
Grenacher, L., Jetzer, Ph., Strassle, M., De Paolis, F. 1999, A\&A, 351, 775

\bibitem[Gurevich et al. 1997]{GZS98}
Gurevich, A.V., Zybin, K.P., Sirota, V.A. 1997, Sov. Phys. Usp., 167, 913

\bibitem[Gyuk \& Holder 1997]{GH97}
Gyuk, G., Holder, G. P. 1997, MNRAS, 297, L44

\bibitem[Gyuk, Dalal and Griest 2000]{GDG00}
Gyuk, G., Dalal, N., Griest, K. 2000, ApJ, 535, 90

\bibitem[Han \& Gould 1996]{HG96}
Han, C., Gould, A. 1996, ApJ, 467, 540

\bibitem[Hu et al. 1994]{Hu94}
Hu, E.M., Haung, J.S., Gilmore, G., Cowie, L.L.  1994, Nature, 371, 493


\bibitem[Jetzer 1998]{J98}
Jetzer, Ph. 1998, {\it Gravitational microlensing}, in {\it Topics on gravitational lensing} (eds. Marino A. A. et al.), Bibliopolis, Napoli

\bibitem[Jetzer et al. 1998]{JSW98}
Jetzer, Ph., Strassle, M., Wandeler, U. 1998, A\&A, 336, 411
 
\bibitem[Kerins 1997]{K97}
Kerins, E.J. 1997, A\&A, 328, 5

\bibitem[Kerins 1998]{K98}
Kerins, E.J. 1998, ApJ, 507, 221

\bibitem[Kerins \& Evans 1998]{KE98}
Kerins, E. J., Evans, N. W. 1998, ApJ, 503, L75

\bibitem[Koopmans \& de Bruyn 2000]{KdB99}
Koopmans, L.V.E., de Bruyn, A.G. 2000, A\&A, 358, 793

\bibitem[Mao \& Paczynski 1996]{MP96}
Mao, S., Paczynski, B. 1996, ApJ, 473, 57

\bibitem[Markovic \& Sommer\,-\,Larsen 1997]{MSL97}
Markovic, D., Sommer\,-\,Larsen, J. 1997, MNRAS, 288, 733


\bibitem[Mendez et al. 1996]{M96}
Mendez, R., Minniti, D., de Marchi, G., Baker, A., Couch, W. 1996, MNRAS, 286, 666


\bibitem[Mer\'a et al. 1998]{MCS98}
M\'era, D., Chabrier, G., Schaeffer, R. 1998, A\&A, 330, 937 

\bibitem[Mollerach \& Roulet 1997]{MR97}
Mollerach, S., Roulet, E. 1997, Phys. Rep., 279, 67

\bibitem[Paczynski 1986]{Pac86}
Paczynski, B. 1986, ApJ, 304, 1

\bibitem[Renault et al. 1997]{R97}
Renault, C. et al. 1997, A\&A, 324, 69

\bibitem[Renault et al. 1998]{R98}
Renault, C. et al. 1998, A\&A, 328, 222

\bibitem[Rix 1996]{R96}
Rix, H.W. 1996, {\it The shape of the dark halos}, in the IAU Symposium no. 169 (eds. Blitz, L., Teuben, P.)

\bibitem[Sackett 1997]{S97}
Sackett, P.D. 1997, ApJ, 483, 103

\bibitem[Sackett 1999]{S99}
Sackett, P.D. 1999, {\it The shape of dark matter halos}, in the ASP Conference Series 182 (eds. Merritt D., Valluri M., Sellwood J.)

\bibitem[Sazhin 1996]{Sazhin}
Sazhin, M.V. 1996, Phys. Lett. A, 219, 199

\bibitem[Wilkinson \& Evans 1999]{WE99}
Wilkinson, M.I., Evans, N.W. 1999, MNRAS, 310, 645

\bibitem[Zhao et al. 1995]{Z95}
Zhao, H.S., Spergel, D.N., Rich, R.M. 1995, ApJ, 440, L13

\end{thebibliography}
\end{document}